\documentclass{article}
\usepackage{amssymb}

\usepackage{epsfig}

\def\be{\begin{equation}}
\def\ee{\end{equation}}
\def\bdm{\begin{displaymath}}
\def\edm{\end{displaymath}}
\def\bea{\begin{eqnarray}}
\def\eea{\end{eqnarray}}
\def\ba{\begin{array}}
\def\ea{\end{array}}

\newcommand{\Bild}[4]{
\begin{figure}[htb]
  \begin{center}
    \leavevmode
    \epsfig{file=#2,height=#1cm}
    \caption{{\small #3}}
    \label{#4}
  \end{center}
\end{figure}}

\begin{document}

\title{\bf $q$-linear approximants: \\
Scaling functions for polygon models}

\date{ \today}

\author{ C.~Richard and A.~J.~Guttmann \\
Department of Mathematics and Statistics\\
The University of Melbourne,
Parkville, Victoria 3052, Australia}

\maketitle

\begin{abstract}

The perimeter and area generating functions of exactly solvable polygon models satisfy
$q$-functional equations, where $q$ is the area variable.
The behaviour in the vicinity of the point where the perimeter generating function 
diverges can often be described by a scaling function.
We develop the method of $q$-linear approximants in order to extract the approximate
scaling behaviour of polygon models when an exact solution is not known.
We test the validity of our method by approximating exactly solvable $q$-linear polygon 
models.
This leads to scaling functions for a number of $q$-linear polygon models, notably
 generalized rectangles, Ferrers diagrams, and stacks.

\end{abstract}

\section{Introduction}

Models of polygons and related combinatorial objects have received considerable 
attention in recent years (for a recent monograph, see \cite{J00}).
They are of interest in physics as models of vesicles or polymer molecules in solution.
The interplay between bulk energy and surface energy in these models gives rise to a
phase transition from an extended phase to a compact, ball-shaped phase \cite{BOP93}.
There have been many studies of combinatorial aspects of these models,
including a general method for deriving the perimeter and area generating function 
of column-convex models \cite{B96}.
Less is known about analytic aspects of the solutions, which are needed
to understand the phase transitions of these models.
Scaling functions which describe the crossover behaviour at critical points have been
computed for a number of polygon models, mostly by indirect methods such as
from a semi-continuous version of the models \cite{PO95, PB95}.
The only direct derivation of scaling behaviour has been for staircase polygons
\cite{P94} by methods of uniform asymptotic expansions.
There is, however, no known general method to obtain scaling functions directly
from functional equations.

This paper presents such a method in the simplest case of a $q$-linear functional 
equation in the perimeter variable.
This class of functional equations is satisfied by rectangles, Ferrers diagrams, 
and stacks \cite{PO95}.
As a step towards the analysis of more complicated classes, we introduce
$q$-linear approximants of first order so as to analyze models which do not obey 
a first-order $q$-linear equation, but which can be well approximated by one.
We will test our method by approximating exactly solvable $q$-linear polygon models 
of generalized rectangles, Ferrers diagrams and stacks.
In particular, we will analyze the model of Ferrers diagrams with a hole and obtain a
differential equation for the scaling function by analysis of the approximants.
We discuss the connection between this new type of approximant and the method of partial
differential approximants \cite{FC82, SF82, RF88, S90}.   
Finally we indicate how our methods can be extended to more general classes of
polygon models.

In a subsequent publication we will consider $q$-quadratic and other non-linear
approximants, which we expect will give good approximations to the scaling
function of as yet unsolved models, such as self-avoiding polygons.

\section{Phase diagrams and scaling functions}

Let us briefly review phase diagrams of $q$-linear polygon models in order to fix
our notation.
(We follow \cite{PO95,PB95,Ow00}).
The perimeter and area generating function of a polygon model is given by
\be
f(x,y,q) = \sum_{r,s,n=1}^\infty f_{r,s,n} x^r y^s q^n = \sum_{n=1}^\infty f_n(x,y) q^n,
\ee
where $f_{r,s,n}$ denotes the number of configurations of area $n$, horizontal perimeter
$r$ and vertical perimeter $s$.
We introduce the area activity $q$, the horizontal perimeter activity $x$, and the
vertical perimeter activity $y$.
The perimeter generating function of the polygon model is given by $f(x,y,1)$.
A phase diagram is the graph of the radius of convergence of $f(x,y,q)$ in
the parameter space $x,y,q$.
Let us consider the isotropic version $f(t,q) := f(t,t,q)$ of the model, with $t$ 
denoting the total perimeter, so that the phase diagram is two-dimensional.
For a typical $q$-linear polygon model such as Ferrers diagrams or stacks, defined below,
the phase diagram is as depicted in Figure \ref{fig:stacphase}. 
\Bild{4}{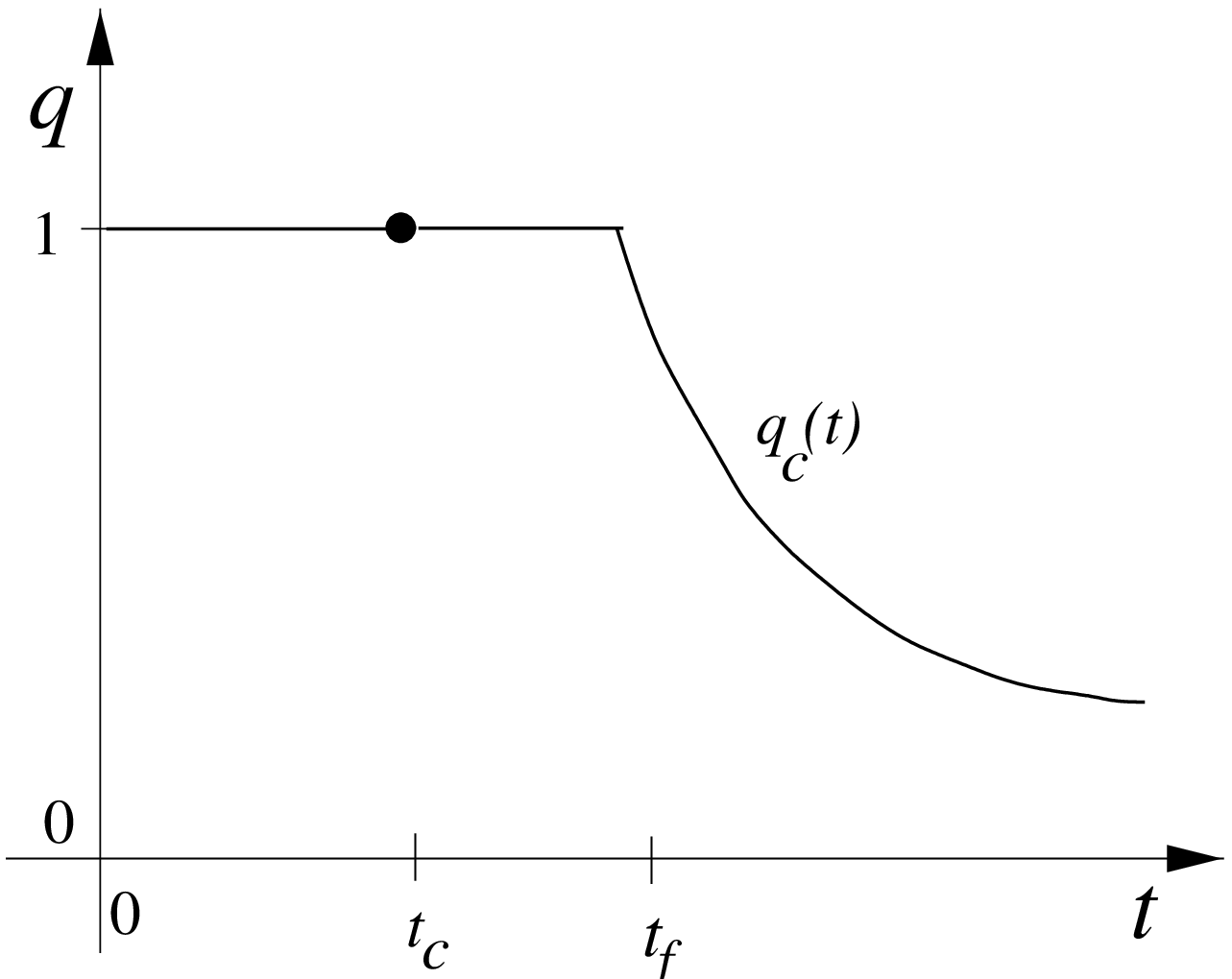}{Phase diagram of a typical $q$-linear polygon
model}{fig:stacphase}
Let us interpret the phase diagram using the grand-canonical ensemble in which we count,
for fixed area, all polygons by perimeter.
The curve $q_c(t)$ where the polygon generating function diverges is related to the 
{\it free energy} per unit area of the ensemble,
\be
-\log q_c(t) = \lim_{n \to \infty} \frac{1}{n} \log f_n(t).
\ee
The phase $q_c(t)<1$ consists of {\it inflated} polygons, whose perimeter grows like 
their area.
The phase $q_c(t)=1$ consists of {\it ball-shaped} polygons, their perimeter growing as
the square-root of their area.
This results in a vanishing free energy for the ensemble.
This behaviour is characteristic of a first order phase transition at the point $t_f$
where both phases meet.
In the ball-shaped phase $q_c(t)=1$, there are contributions to the {\it boundary} free 
energy, however\footnote{
The boundary free energy is defined as the limit
$f_b(t)= \lim_{n \to \infty} \frac{1}{\sqrt{n}} \log f_n(t)$.
}.
Let us denote by $t_c$ the point where the perimeter generating function diverges.
At this point a phase transition in the boundary free energy occurs:
For $t<t_c$, the contributions to the boundary free energy are given by polygons of finite
size, whereas for $t>t_c$ the contributions to the boundary free energy derive from
polygons of infinite size.

In the remainder of this paper we will concentrate on the critical behaviour about the 
point $t_c$ where the perimeter generating function diverges.
This point is the natural one to look at from the perspective of power series
approximations, which we employ.
Moreover, for rectangles and more complicated models such as self-avoiding polygons, the 
distinction between the two phase transitions is irrelevant since $t_c$ and $t_f$ coincide for 
these models. 

To describe the singular behaviour about $t_c$ in more detail,
consider $f(t,q)$ for $t$ fixed, as $q$ approaches unity.
For $q$-linear polygon models, $q=1$ is a point of an essential singularity in the
generating function:
For $t<t_c$, $f$ converges to a finite limit.
If $t=t_c$, $f$ has a power-law divergence with an exponent generally different
from that of the perimeter generating function.
If $t>t_c$, $f$ diverges with an essential singularity.
In many cases, the crossover between these types of critical behaviour can be described 
by a scaling function $\bar{P}(\bar{s})$ of combined argument 
$\bar{s}= (t_c-t)(1-q)^{-\phi}$,
\be
f(t,q) \sim \frac{1}{(1-q)^\theta} \bar{P}\left( \frac{t_c-t}{(1-q)^\phi}\right) 
\qquad \left( ( t,q) \to (t_c^-,1^-) \right). \label{eqn:scalingfctn}
\ee
The asymptotic behaviour of the scaling function at infinity is related to the behaviour
of $f(t,q)$ for $t<t_c$.
To see this, assume that $f(t,q)$ admits an asymptotic expansion of the form
\be
f(t,q) = \sum_{n=0}^\infty f_n(t) (1-q)^n \qquad (t<t_c)
\ee
about $q=1$, where the leading contributions of the coefficients $f_n(t)$ are given by
\be
f_n(t) = \frac{p_n}{(t_c-t)^{\gamma_n}} + {\cal O} \left( (t_c -
t)^{-\gamma_n+1}\right),
\ee
as $t$ approaches $t_c$%
\footnote{The coefficients may have a different asymptotic form, see rectangles at 
$y=1$ below.}.
For $q$-linear polygon models, the asymptotic expansion can be computed recursively from
the defining functional equation.
It can be inferred from (\ref{eqn:scalingfctn}) that the existence of a scaling function
implies the restriction
\be
\gamma_n = \frac{\theta}{\phi} + \frac{n}{\phi} \label{eq:gamn}
\ee
on the exponents $\gamma_n$.
Moreover, it can be seen that the numbers $p_n$ are the coefficients in the asymptotic
expansion of the scaling function
\be
\bar{P}(\bar{s}) = \sum_{n=0}^\infty p_n \, \bar{s}^{-\gamma_n}.
\ee
We assume that the scaling function $\bar{P}(\bar{s})$ is regular at the origin.
(This assumption is not always fulfilled.
The simplest counterexample is the model of rectangles in its isotropic version.)
In this situation, the behaviour of $f$ at $t=t_c$ is given by
\be
f(t_c,q) \sim \frac{\bar{P}(0)}{(1-q)^\theta} \qquad (q \to 1^-).
\ee
The exponents $\theta$ and $\gamma_0$ are called {\it critical exponents} of the 
model.
$\theta$ describes the behaviour of $f(t_c,q)$ about $q=1$, whereas 
$\gamma_0$ describes the power-law behaviour of the perimeter generating function 
about $t_c$. 
The exponent $\phi$ is called the {\it crossover exponent} and relates the two critical
exponents, see (\ref{eq:gamn}).

\section{$q$-linear polygon models}

We call a polygon model $q$-linear of $N$th order if its generating function satisfies 
a $q$-linear functional equation%
\footnote{
Properties of solutions of $q$-linear functional equations if $q<1$ have been studied
in \cite{A31}, for recent results see \cite{APP98}.}
\be
G(t,q) = \sum_{k=1}^N a_k(t,q) G(q^kt,q) + b(t,q), \label{form:leq}
\ee
where $a_k(t,q)$ and $b(t,q)$ are rational functions in $t$ and $q$.
Here, $t$ may denote the total or horizontal or vertical perimeter.
Explicit realizations include rectangles, Ferrers diagrams, and 
stacks \cite{PO95}.
They all satisfy a $q$-linear functional equation of first order in the horizontal perimeter
activity $x$.

We may construct new polygon models from given ones by allowing for decorations of the
polygons.
In this way we may obtain models of polygons with holes, of coloured polygons, or the like.
Decorations may be interpreted in physical terms as allowing for a refined structure of
polygons, which may be a better approximation to vesicles than undecorated polygons.
These models have a natural interpretation in terms of random tilings of polygonal shape 
\cite{RHHB98}.

The question arises as to which decorations lead to models which continue
to satisfy $q$-linear functional equations.
This is the case for models whose generating function can be obtained by application of a
linear differential operator (w.r.t. $x$, $y$, and $q$) to the generating function of the
undecorated model.
For example, the model of Ferrers diagrams with a 1-hole, defined in the appendix, satisfies a
$q$-linear functional equation of order 3.
Let us now focus on the simplest class of decorated polygon models which satisfy a $q$-linear
functional equation of first order.
To this end, consider first decorated rectangles of unit height.
We denote the generating function of this model by $b_1(x,y,q)$.
For example, the generating function of a rectangle with exactly $k$ black unit squares is
given by
\be
b_1(x,y,q) = \frac{y (qx)^k}{(1-qx)^{k+1}},
\ee 
while the generating function of undecorated rectangles is given by 
$b_1(x,y,q)=yqx/(1-qx)$.
Let us now consider the polygon model of rectangles with a decorated top layer.
As indicated in Fig. \ref{fig:functional}, 
\Bild{7}{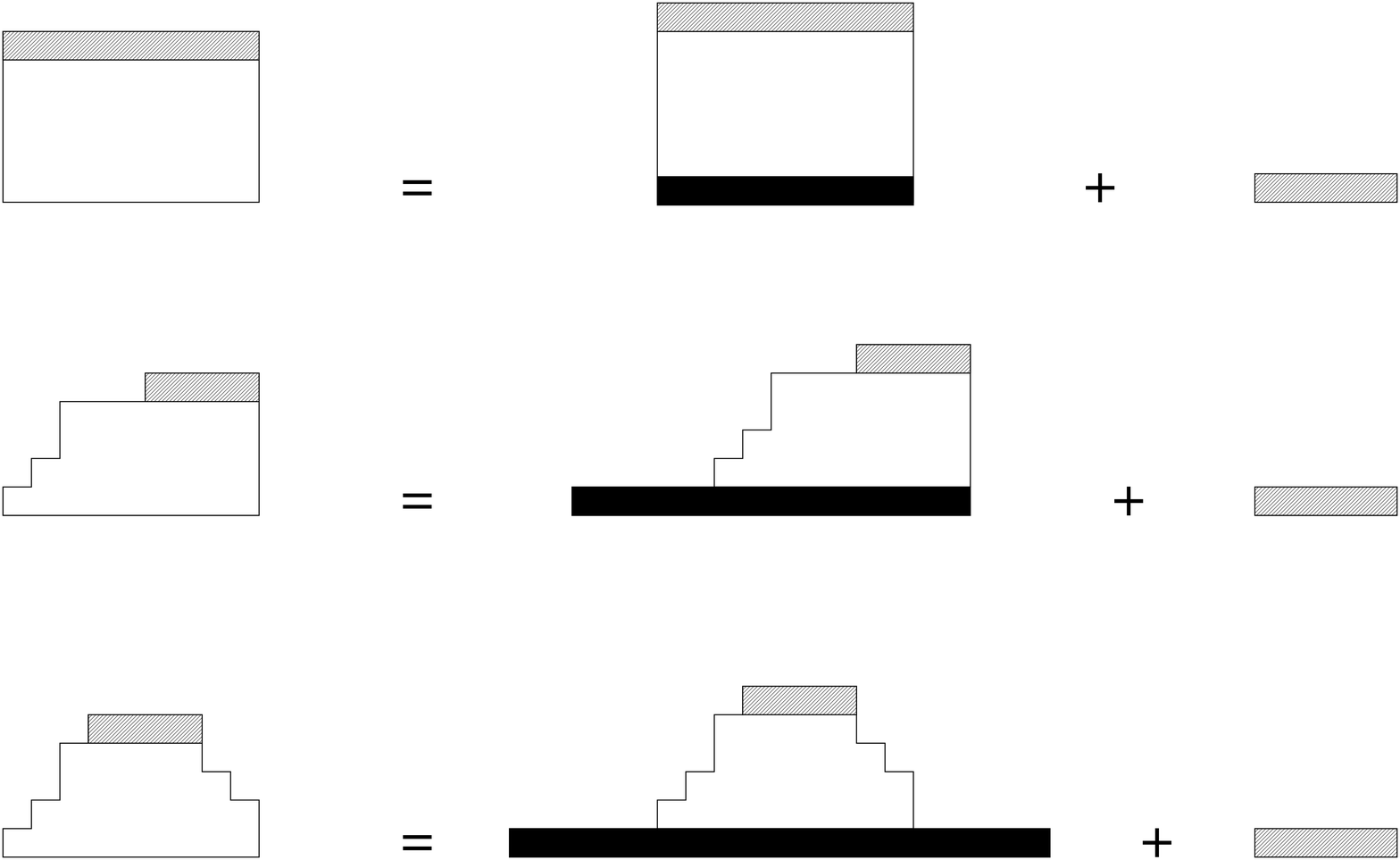}{Graphical representation of the functional
equations for rectangles, Ferrers diagrams, and stacks with a decorated top 
layer.}{fig:functional}
we can consider all rectangles of height $m+1$ as
being obtained from rectangles of height $m$ by adding a row of squares below the bottom layer.
This construction misses out all decorated rectangles of height $1$.
A similar construction can be applied to Ferrers diagrams and stacks with a decorated top layer.
The figure indicates that the models defined above satisfy a $q$-linear functional 
equation of first order
\be
G_s(x,y,q) = \frac{y}{(1-qx)^s}G_s(qx, y, q) + b_1(x,y,q), \label{form:funceq}
\ee
where $s=0,1,2$ denotes rectangles, Ferrers diagrams, and stacks respectively, and
$b_1(x,y,q)$ is the generating function of decorated rectangles of unit height.
Equation (\ref{form:funceq}) can be iterated to give a closed form for the
area and perimeter generating function
\be
G_s(x,y,q) = \sum_{n=1}^\infty \frac{y^{n-1} b_1(q^n x,y,q)}{(qx;q)_{n-1}^s}, 
\label{form:linchain}
\ee
where $(t;q)_n = \prod_{k=0}^{n-1}(1-q^k t)$ denotes the $q$-product.
The perimeter generating function can be obtained from the functional equation by setting
$q=1$ and solving for $G_s(x,y,1)$,
\be
G_s(x,y,1) = \frac{(1-x)^s}{(1-x)^s-y} b_1(x,y,1).
\ee
The models of rectangles, Ferrers diagrams and stacks are recovered as special cases
where the trivial decoration is chosen.

\section{Scaling functions via dominant balance}

We now discuss how to derive scaling exponents and scaling functions from
$q$-linear functional equations about the critical point where the perimeter generating
function diverges.
Our technique relies on the method of dominant balance \cite{BO78}, which has previously 
been applied to the derivation of scaling equations for a number of semi-continuous models 
\cite{PB95,PO95}.
Our approach lies in deriving scaling functions for the discrete models by manipulating
the $q$-functional equation directly.
It consists of three steps:
Firstly, the critical point is shifted to the origin by a change of variables.
Then a scaling variable is introduced, and a consistent set of scaling exponents is 
sought.
Finally, the resulting differential or difference equation is solved.

Consider the $q$-linear functional equation of first order
\be
a_0(x,q) f(x,q) - a_1(x,q)f(qx,q) - b(x,q)=0. \label{form:dbeq}
\ee
Bearing later applications in mind, we restrict $a_0(x,q), a_1(x,q), b(x,q)$ to
polynomials in $x$ and $q$.
For readability, we suppress all subsequent dependencies on $q$.
As a first step, assume that the critical point is at $x=x_c$ and $q=1$.
We expand the functional equation about the critical point.
To this end we introduce small variables $\epsilon=1-q$ and $s=x_c-x$,
and define $P(s)=f(x_c-s)$.
In these variables, the functional equation reads
\be
\left( a_0 - a_1\right)(x_c-s) P(s) 
- a_1(x_c-s) \sum_{n=1}^\infty \frac{(\epsilon(x_c-s))^n}{n!} \frac{d^n}{ds^n} P(s) - b(x_c-s)
=0. \label{form:shifted}
\ee
Second, we introduce scaled quantities
\be
s = \epsilon^\phi \bar{s}, \qquad P = \epsilon^{-\theta} \bar{P},
\ee
and write equation (\ref{form:shifted}) in terms of $\bar{s}$ and $\bar{P}$.
These quantities are just the scaling variable of combined argument and the scaling function 
(\ref{eqn:scalingfctn}).

Here, $\phi$ is assumed to be positive.
This leads to additional factors of $\epsilon$ in each summand.
The scaling equation results from taking only the terms with {\it smallest} exponents in
$\epsilon$, for suitable choices of the exponents $\theta$ and $\phi$.
 
Let us analyze the contribution from the $n$-th order in the expansion of $f(qx)$.
Scaling leads to exponents of the form $n(1-\phi)$.
This implies the constraint $\phi \le 1$, since other values lead to arbitrarily small
exponents with increasing $n$.
If $\phi < 1$, only the first order contributes, leading to 
$x_c \frac{d}{d\bar{s}} \bar{P}(\bar{s}) $.
If $\phi=1$, higher orders cannot be ignored, and the dominant part of the sum equals
$\bar{P}(\bar{s} + x_c) - \bar{P}(\bar{s})$.
This results in a differential or difference equation for the scaling function.
We will present typical examples of both kinds below.

We concentrate on critical points given by the smallest pole of the perimeter generating
function, that is at the smallest positive $x_c$ satisfying
\be
a_1(x_c,1) = a_0(x_c,1). 
\ee
For polygon models, the coefficients of the perimeter generating
function are all positive. This generating function is obtained by setting
$q = 1$ in (\ref{form:shifted}), and hence implies that $b(x)/(a_0(x)-a_1(x))$ has nonnegative 
Taylor-coefficients.
This implies in particular that $b(x_c,1)$ is non-zero.
Therefore, the leading contribution of $b(x)$ in the scaling limit is of order
$\epsilon^0$.
The contributions from $a_0(x)-a_1(x)$ and from $a_1(x)$ have to be analyzed for each model
separately.
To obtain a nontrivial scaling equation, we demand that we get contributions from each
of the three terms in (\ref{form:shifted}).
This means that all three exponents in $\epsilon$ have to be equal. 
We thus arrive at a set of equations determining $\theta$ and $\phi$,
\be
\ldots - \theta = \ldots - \theta + (1-\phi) = 0,
\ee
where $\ldots$ denotes contributions from $a_0(x)-a_1(x)$ and $a_1(x)$, respectively.
We discuss particular examples below.
We finally mention that a scaling analysis of the $q$-linear equation of order $N$ can 
be carried out by the same method.
If $\phi<1$, this results in a differential equation, as found above for the case $N = 1.$

\subsection{Ferrers diagrams}

Ferrers diagrams satisfy a $q$-linear functional equation with
\begin{eqnarray}
a_0(x,q)&=& 1-qx,\\
a_1(x,q)&=& y, \nonumber \\
b(x,q)&=& y q x. \nonumber
\end{eqnarray}
The perimeter generating function diverges at $x_c=1-y$.
$a_1(x,q)$ and $b(x,q)$ are nonzero at the critical point, whereas
$a_0(x,q)-a_1(x,q)$ 
vanishes linearly in $s=x_c-x$.
The leading contributions in the three terms of equation (\ref{form:shifted}) give
\be
\phi - \theta = -\theta + (1-\phi) = 0.
\ee
This leads to exponents 
\be
\theta= \frac{1}{2}, \qquad \phi=\frac{1}{2},
\ee
and to the differential equation
\be
\frac{1}{y(1-y)} {\bar s} {\bar P}({\bar s}) =
\frac{d}{d{\bar s}} {\bar P}({\bar s}) + 1.
\ee
The solution, which is uniquely determined if we demand power-law behaviour as
$\bar{s} \to \infty$, is given by a complementary error-function
\be
\bar{P}(\bar{s}) = \sqrt{\frac{\pi a}{2}} \, \mbox{erfc}\left(
\frac{\bar{s}}{\sqrt{2a}}\right) \exp\left(\frac{\bar{s}^2}{2a} \right),
\ee
where $a=y(1-y)$.
The asymptotic behaviour is given by
\begin{eqnarray}
\bar{P}(\bar{s})  &\sim& \frac{a}{\bar{s}} \qquad (\bar{s} \to +\infty)  \\
\bar{P}(\bar{s})  &\sim&  \sqrt{2\pi a} \, \exp\left(\frac{\bar{s}^2}{2a} \right) 
\qquad (\bar{s} \to -\infty)  \nonumber \\
\bar{P}(0) &=& \sqrt{\frac{\pi a}{2}}. \nonumber
\end{eqnarray}
This is in agreement with both the scaling function found for
the semi-continuous model and with 
the asymptotic behaviour computed previously \cite{PO95}.

\subsection{Decorated rectangles}

We next consider the model of rectangles with a decorated top layer.
The decoration consists of exactly $k$ black squares.
The generating function of the top layer is
\be
b_1(x,y,q) = \frac{y (qx)^k}{(1-qx)^{k+1}},
\ee 
As shown above, the model satisfies a $q$-linear functional equation with
\begin{eqnarray}
a_0(x,q)&=& (1-qx)^{k+1},\label{eq:rectchain}\\
a_1(x,q)&=& y(1-q x)^{k+1}, \nonumber \\
b(x,q)&=& y (q x)^k. \nonumber
\end{eqnarray}
The perimeter generating function has a pole of order $k+1$ at $x_c=1$.
In contrast with the model of Ferrers diagrams, this point coincides with the point $x_f$
where the first order phase transition in the free energy occurs.
The case $k=0$ is closely related to rectangles\footnote{For rectangles, 
$b_1(x,y,q)=yqx/(1-qx).$}: 
Comparison of (\ref{eq:rectchain}) with the functional equation for rectangles shows
that both models obey the same equations about the critical point and hence have the same
scaling functions. 

The leading terms in the three summands of equation (\ref{form:shifted}) give
\be
(k+1)\phi - \theta = (k+1)\phi - \theta + (1-\phi) = 0.
\ee
This leads to exponents 
\be
\theta = k+1, \qquad \phi=1,
\ee
and we get the difference equation
\be
(1+ \bar{s})^{k+1} \left( \bar{P}_k(\bar{s}) - y \bar{P}_k(\bar{s}+1) \right) -y =0.
\ee
In order to obtain the correct scaling function from this recursion, we have to force
the asymptotic behaviour as $\bar{s} \to \infty$ to be of power-law type.
This fixes the constant term, and for $y<1$ we arrive at the Lerch functions
\be
\bar{P}_k(\bar{s}) = \sum_{n=1}^\infty \frac{y^n}{(n+\bar{s} )^{k+1}}.
\ee
The asymptotic behaviour is given by
\begin{eqnarray}
\bar{P}_k(\bar{s})  &\sim& \frac{y}{1-y}\frac{1}{\bar{s}^{k+1}} 
\qquad (\bar{s} \to +\infty)  \\
\bar{P}_k(0) &=& \sum_{n=1}^\infty \frac{y^n}{n^{k+1}} . \nonumber
\end{eqnarray}
The case $k=0$ (rectangles) is in agreement with both the scaling function 
of the semi-continuous model and the asymptotics computed in \cite{PO95}.

The case $y=1$ is different.
The perimeter generating function diverges for all values of $x$, but it is
possible to obtain scaling behaviour about $x_c=1$, as $q$ approaches unity.  
For $k>0$, the solution of the difference equation is given by
\be
\bar{P}_k(\bar{s}) = \frac{(-1)^{k+1}}{k!}\Psi_k(1+\bar{s})
\ee
where $\Psi_k(x)$ denotes the $k$-th derivative of the $\Psi$-function
$\Psi(x)=\partial_x \log\Gamma(x)$.
The asymptotic behaviour is given by
\begin{eqnarray}
\bar{P}_k(\bar{s})  &\sim& \frac{1}{k}\frac{1}{\bar{s}^k} 
\qquad (\bar{s} \to +\infty)  \\
\bar{P}_k(0) &=& \zeta(k+1), \nonumber
\end{eqnarray}
where $\zeta(k)=\sum_{n=1}^\infty n^{-k}$ denotes the Riemann zeta function. 

If $k=0$, the scaling function diverges logarithmically as 
$\bar{s} \to \infty$.
In order to obtain a well-defined crossover behaviour, we compensate for this by
the addition of a logarithmic term,
\be
\bar{P}_0(\bar{s}) = -\Psi(1+\bar{s}) -\log\epsilon. \label{eq:rect0}
\ee
The asymptotic behaviour then follows as
\begin{eqnarray}
\bar{P}_0(\bar{s})  &\sim& -\log (\bar{s}\epsilon) = - \log s
\qquad (\bar{s} \to +\infty)  \\
\bar{P}_0(0) &=& -\log \epsilon + \gamma, \nonumber
\end{eqnarray}
where $\gamma=-\int_0^\infty e^{-t} \ln t \, dt \approx 0.57721$ denotes Euler's constant.
This gives the (non-uniform) asymptotic behaviour first computed in \cite{PO95}.
The scaling function describing this behaviour was not previously known, however.

\section{$q$-linear approximants}

We now consider the more general situation where we cannot obtain the generating
function, but only a finite number of terms thereof.

\subsection{The method}
The basic idea of $q$-linear approximants (of first order) is to fit a $q$-linear 
functional equation to a function $f(t,q)$,
\be
a_0(t,q) f(t,q) = a_1(t,q) f(qt,q) + b(t,q), \label{form:appeq}
\ee
such that (\ref{form:appeq}) is {\em exact} up to a given degree in the variables.
Here, we restrict $a_0(t,q),$ $a_1(t,q),$ and $b(t,q)$ to be polynomials in $t$ and $q$ of degree
$n_t$ and $n_q$, say.
The coefficients of the polynomials can be found by solving the system of linear 
equations deriving from the expansion of (\ref{form:appeq}) in its two variables.
This process is not unique, since there are many choices of sets of linear equations.
Moreover, (\ref{form:appeq}) could be expanded about points other than the origin,
resulting in approximants accurate about these points.
Our approach is to demand that $q$-linear approximants shall reduce to 
{\it linear approximants} with polynomials of order $n_t$, as $q$ approaches unity, which
corresponds to an approximation of the perimeter generating function by rational
functions. Thus such approximants are only likely to be good if the perimeter
generating function is dominated by a pole.
We therefore expand  (\ref{form:appeq}) about $t=0$ and $q=1$, taking into account only
terms up to a fixed order $N_t(n_t)$ in $t$.
In order to obtain the desired limit, $N_t(n_t)$ has to be chosen large enough.
We order the resulting terms by increasing powers in $1-q$ and, for each power,
by increasing powers in $t$.
For given $N_t(n_t)$, we compute $q$-linear approximants with polynomials of order $n_t$
in $t$ and of order $n_q=0,1,\ldots$
For given $N_t(n_t)$, the highest obtainable order $n_q$ depends on the number of 
equations deriving from (\ref{form:appeq}).
To fix the multiplicative constant, we require $a_0(0,0)=1+a_1(0,0)$.
Information about the scaling function is obtained by applying the method of
dominant balance to the approximants, as described above.

In order to test the method of $q$-approximants, we will apply the method to a number of
exactly solvable $q$-linear polygon models, which are mainly isotropic versions of the
models defined above.
These models have a rational perimeter generating function and can be shown to obey a $q$-linear 
equation of higher than first order.
Therefore, $q$-linear approximants (of first order) should give the correct differential
equation for the scaling function.
This can be checked by computing the leading coefficients in the asymptotic expansion of
the model, using the anisotropic functional equation directly.
If the functional equation of the isotropic model is known, the scaling function can
alternatively be obtained by applying the method of dominant balance.

We will illustrate the method of $q$-approximants by deriving the scaling behaviour of
the $q$-linear models, defined above in the isotropic case, which do not obey a
$q$-linear functional equation of first order.

\subsection{Stacks}

The generating function $f(t,q)$ of stacks with equal horizontal and vertical
perimeter activity $x=y=t$ satisfies the $q$-linear functional equation 
of order 2
\be
a_0(t,q) f(t,q) = a_1(t,q) f(qt,q) +a_2(t,q) f(q^2t,q) + b(t,q),
\ee
where the polynomials $a_0(t,q)$, $a_1(t,q)$, $a_2(t,q)$ and $b(t,q)$ are given by
\begin{eqnarray}
a_0(t,q)&=& (1-q^2t)^2 (1-qt)^3, \label{eq:stacks}\\
a_1(t,q)&=& t^2(1+q)(1-q^2t)^2, \nonumber \\
a_2(t,q)&=& -q^3t^4, \nonumber \\
b(t,q)&=& -qt^2(1-q^2t)(q^4t^3-q^3t^2+q^2t+qt-1). \nonumber
\end{eqnarray}
We found this relation by computing $q$-linear approximants of second order to stacks.
It is possible to interpret this relation combinatorially \cite{BM01}.
The perimeter generating function $f(t,1)$ diverges at $t_c=(3-\sqrt5)/2$ with a simple pole.
The method of dominant balance can be applied to compute the scaling function about $t_c$.
It is of the same form as the scaling function for Ferrers diagrams, which corresponds
to the observation made for the semi-continuous models \cite{PO95}.
We used stacks in order to test the method of $q$-linear approximants. 
Their rational perimeter generating function is obtained by approximants 
of cubic order ($n_t$=3).
A scaling analysis of cubic approximants at $q=1$ and $t_c=(3-\sqrt5)/2$ 
yields the correct type of differential equation for the scaling functions for each 
approximant, with generally incorrect coefficients.
The accuracy of approximation increases with the degree $n_q$ in $(1-q)$ of the 
polynomials.
For $N_t(3) > 6$ and $n_q\ge1$, the approximants yield the {\em correct} scaling 
equation.
This result is robust against increasing the order $n_t$.

\subsection{Decorated Ferrers diagrams}

We introduce Ferrers diagrams with a decorated top layer: 
Consider a decoration consisting of exactly $k$ black squares.
Approximants indicate that the scaling function is the same as for (pure) Ferrers 
diagrams. 
We checked this for $k=0,1,\ldots,5$.

Instead of considering more complicated decorations of the top layer, we will now consider 
Ferrers diagrams with a decorated bottom layer. 
Let us approximate the model with a decoration consisting of exactly two black squares
which may be placed anywhere in the bottom layer.
The model has the generating function
\be
G(x,y,q) = \sum_{n=2}^\infty \frac{n(n-1)}{2} \frac{y (qx)^n}{(qy;q)_n},
\label{form:ferrch}
\ee
where $(t;q)_n = \prod_{k=0}^{n-1}(1-q^k t)$ denotes the $q$-product.
The perimeter generating function of the model is
\be
G(x,y,1) = \frac{x y^2(1-x)}{(1-x-y)^3}.
\ee
We consider the isotropic case where $x=y=t$.
The critical point is $t_c=1/2$.
Since the perimeter generating function is rational with numerator and denominator
polynomials of order 4 and 3, we use $q$-linear approximants with $n_t=4$ and 
$N_t=20$.
For $n_q>2$, the dominant terms of the approximants give the differential equation  
\be
\left( {\bar s}^3 + \frac{3}{16} {\bar s} \right)  {\bar P}({\bar s}) -
\left( \frac{{\bar s}^2}{16}+ \frac{1}{256} \right)
\frac{d}{d{\bar s}} {\bar P}({\bar s})
- \frac{1}{128} =0
\ee
with exponents
\be
\theta=\frac{3}{2}, \qquad \phi=\frac{1}{2}.
\ee
This result is robust against varying the values of $n_q$ and $n_t$ in the approximation.
We checked this up to $n_q=8$ and also for increasing values of $n_q$ at $n_t=5$ and 
$n_t=6$.
The above equation leads to the scaling function
\be
 {\bar P}({\bar s}) = \left( \int_{\bar s}^\infty \frac{e^{-8 t^2}}{(1+16
t^2)^2} \, d t  \right) 2 e^{8 {\bar s}^2} (1+16 {\bar s}^2).
\ee
The asymptotic behaviour is given by
\begin{eqnarray}
\bar{P}(\bar{s})  &\sim& \frac{2^{-7}}{\bar{s}^3} \qquad (\bar{s} \to +\infty)  \\
\bar{P}(\bar{s})  &\sim&  2 \bar{P}(0) \, e^{8 {\bar s}^2} {\bar s}^2
\qquad (\bar{s} \to -\infty) \nonumber \\
\bar{P}(0) &=& \frac{\sqrt{2\pi}}{8} \nonumber.
\end{eqnarray}
Using the methods described at the end of the appendix, it can be shown that the isotropic model satisfies a
$q$-linear equation of third order which can be used to test that the scaling function
obtained by $q$-linear approximants is correct.

\subsection{Ferrers diagrams with a 1-hole}

Ferrers diagrams with a 1-hole are defined in the Appendix.
We again consider the isotropic model where $t=x=y$.
Since the perimeter generating function is rational with numerator and denominator
polynomials of order 6 and 5, we use $q$-linear approximants with $n_t=6$ and $N_t =
20$.
For $n_q>2$, the dominant terms of the approximants give the differential equation  
\be
\left( {\bar s}^3 + \frac{3}{16} {\bar s} \right)  {\bar P}({\bar s}) -
\left( \frac{{\bar s}^2}{16}+ \frac{1}{256} \right)
\frac{d}{d{\bar s}} {\bar P}({\bar s})
- \frac{1}{128} =0
\ee
with exponents
\be
\theta=\frac{3}{2}, \qquad \phi=\frac{1}{2}.
\ee
This is the same equation as that for decorated Ferrers diagrams.
This result is robust against varying the values of $n_q$ and $n_t$ in the approximation.
We checked this up to $n_q=8$ and also by increasing values of $n_q$ at $n_t=7$ and 
$n_t=8$.

\section{Comparison with partial differential approximants}

There is an existing approximating method designed to compute critical
exponents and scaling functions about multi-singular points, known as
the method of partial differential approximants (p.d.a.), due to Fisher 
and co-workers \cite{FC82, SF82, RF88, S90}. 
We will briefly explain the method and compare it to our approach.

The basic idea of p.d.a. derives from the observation that a scaling function
\be
f(t,q) = \frac{1}{(q_c-q)^\theta} \bar{P} \left( \frac{t_c-t}{(q_c-q)^\phi} \right) + f_0
\label{form:pda}
\ee
obeys the first-order partial differential equation 
\be
\theta f(t,q) + f_0 = \phi (t_c-t) \frac{d}{dt} f(t,q) + (q_c-q) \frac{d}{dq} f(t,q).
\label{form:pde}
\ee
Therefore,  p.d.a. of the form
\be
a(t,q) f(t,q) + b(t,q) = c(t,q) \frac{d}{dt} f(t,q) + d(t,q) \frac{d}{dq} f(t,q),
\label{form:pdeq}
\ee
where $a(t,q)$, $b(t,q)$, $c(t,q)$ and $d(t,q)$ are polynomials in $t$ and $q$,
may serve to detect possible scaling behaviour about {\it multi-critical points}
$(t_c,q_c)$ defined by the simultaneous vanishing of $c$ and $d$,
\be
c(t_c,q_c) = d(t_c,q_c) = 0.
\ee
The critical exponents $\theta$ and $\phi$ can then be read off as lowest order
coefficients in the expansion of the approximating polynomials about the multi-critical
point.
Numerical methods can be used to determine subsequent terms in the Taylor-expansion of
the scaling function.
It has been shown \cite{RF88} that numerical integration works if the crossover
exponent $\phi$ is restricted to $1/2<\phi<2$.

In contrast to this general setup, $q$-linear approximants are only suited to the 
detection of critical behaviour for models whose scaling function obeys a linear 
differential or difference equation of first order.
Equivalently, $q$-linear approximants can be used to test whether a scaling function
obeys an equation of the above type:
If it does not, the approximants are likely to fail to converge.
If it does, the approximants will converge and give the underlying differential or
difference equation.
This is then more specific information than can be gained from the p.d.a. approach,
though the p.d.a. approach currently approximates a broader range of scaling
behaviour. In subsequent work we will extend to non-linear approximants, which
should combine the generality of the p.d.a. approach with the specificity of
information obtainable by the $q$-approximants.

We conclude with remarks about Ferrers diagrams and rectangles.
Since Ferrers diagrams have a crossover exponent $\phi=1/2$, numerical integration using
p.d.a. \cite{RF88}  to obtain the coefficients of the scaling function is unlikely to
converge.
The model of rectangles (\ref{eq:rect0}) does not obey a scaling law of type 
(\ref{form:pda}).
It can be shown that the scaling function gives {\it quadratic} prefactors for the
derivatives in (\ref{form:pde}).
Therefore, the p.d.a. method cannot be successfully applied here in its standard form.
For rectangles with the random chain considered above, p.d.a. may provide good
estimates.  

In summary, we would expect that the method reported here is likely to be better than 
p.d.a. for those systems whose scaling function is described by, or well approximated by, 
a linear differential or difference equation of first order, while the p.d.a. method 
might be expected to better approximate those systems that do not. As we extend the 
method reported here to higher order $q$-functional equations, and possibly other types 
of functional equation as well, we would expect these new $q$-approximants to more 
appropriately represent a correspondingly larger class of systems. We emphasise that 
these are remarks of a general nature, and not the result of rigorous numerical 
comparisons, which we have not carried out.

\section{Conclusion}

We have developed techniques to obtain scaling functions for $q$-linear polygon models 
about the point where the perimeter generating function diverges, using the method of 
dominant balance.
This led to scaling functions for a number of $q$-linear polygons models generalizing
rectangles, Ferrers diagrams and stacks.
The question arises as to what extent can the scaling behaviour for these models 
be obtained by direct methods?
The most direct approach is to approximate the perimeter and area generating function
by their Euler-Maclaurin sum and to estimate the resulting integral by methods of uniform
asymptotic expansions \cite{W89}, in the spirit of \cite{P94}.
The authors are, however, not aware of standard methods to do this, apart from 
Bleistein's method \cite{O74}, which can be used to analyze Ferrers diagrams.

We introduced the method of $q$-linear approximants to obtain scaling functions of 
models where an exact $q$-linear functional equation of first order does not exist.
The method yields the correct scaling functions for exactly solvable models which can be 
described by $q$-linear functional equations of higher than first order, such as for 
decorated $q$-linear models.
We claim that $q$-approximants will be appropriate 
for the analysis of the scaling behaviour of statistical models whose scaling function may be
well described by a difference or differential equation of first order.

The method of dominant balance can also be applied to obtain differential equations for
scaling functions from $q$-functional equations different from $q$-linear.
For example, it is possible to derive the differential equation for 
the scaling function of the model of staircase polygons \cite{PB95}, which belongs to 
the $q$-quadratic class.  
This indicates that the idea of $q$-linear approximants can be generalized to more complex 
classes of polygon models satisfying $q$-algebraic functional equations.
This leads to $q$-algebraic approximants.
Even the $q$-quadratic class is interesting to analyze, since there are solvable models
where the scaling behaviour is not known (such as staircase polygons with a hole; see
also \cite{J001} for an example), and more interestingly it
can be used to approximate the generating function for self-avoiding polygons, 
which displays a square-root divergence in its perimeter generating function.
We are developing our method in that direction.

\section*{Acknowledgements}

We thank M. Bousquet-M\'elou for a number of comments on the manuscript and A. Owzcarek
for stimulating discussions.
Financial support from the German Science Foundation (DFG)
and from the Australian Research Council (ARC) is gratefully acknowledged.

\section*{Appendix: Ferrers diagrams with a 1-hole}

We define the polygon model of Ferrers diagrams with a 1-hole and give a closed
form for the perimeter generating function and for the perimeter and area
generating function.

The model of Ferrers diagrams with a 1-hole consists of all Ferrers diagrams where a
unit square is removed from the interior.
For a typical Ferrers diagram, this is graphically depicted in Fig.
\ref{fig:Ferrerhole}.
\Bild{5}{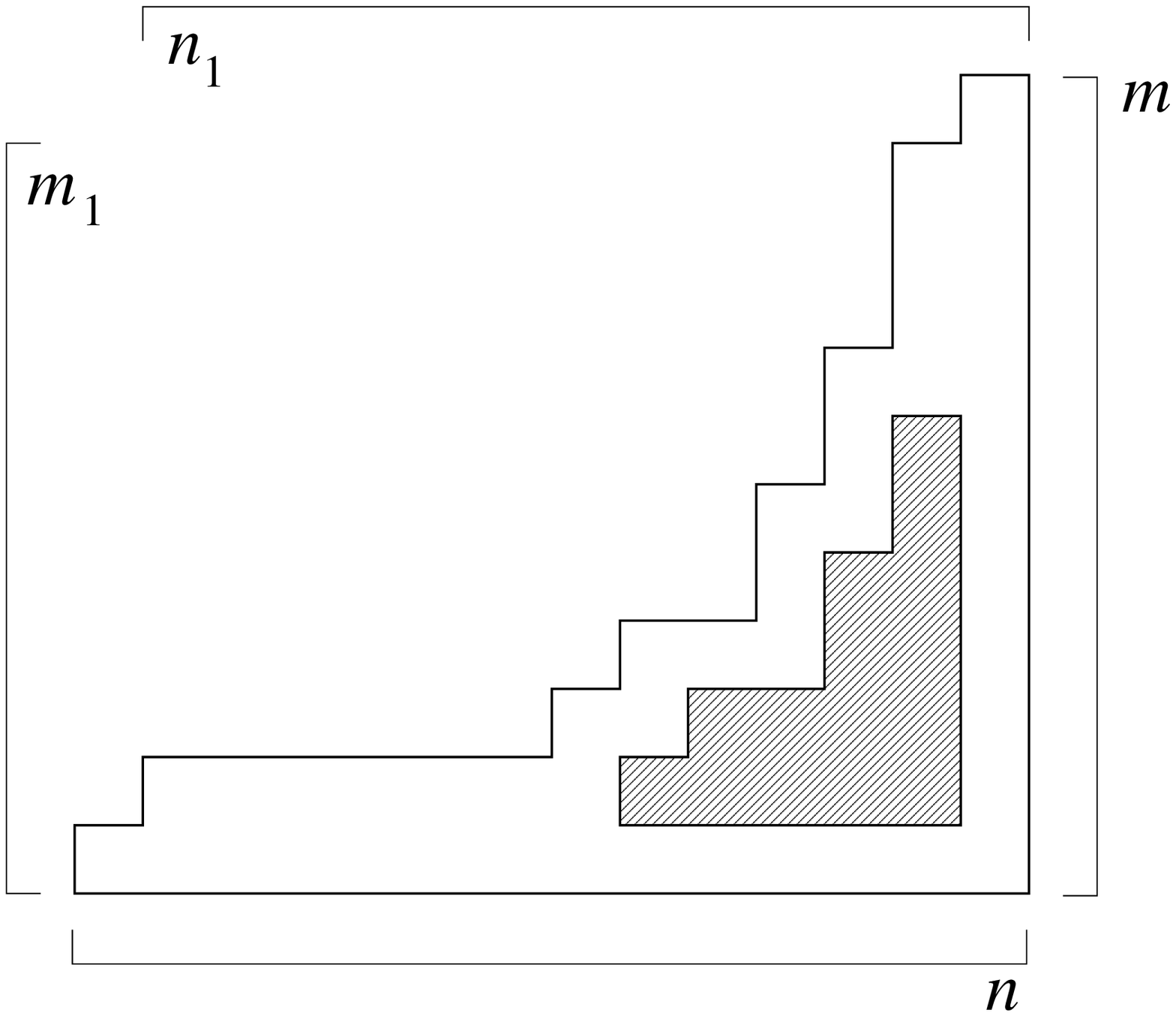}{A typical Ferrers diagram. A hole may occur anywhere in the
shaded region.}{fig:Ferrerhole}
If we denote the last column height by $m$, the last-but-one column height by $m_1$, the
bottom row length by $n$ and the second row length by $n_1$, the number of
possible sites for a hole $h(n,m,n_1,m_1)$ is given by
\be
h(n,m,n_1,m_1) = \mbox{ area } - n -m -n_1-m_1 +4.
\label{form:hole}
\ee
This translates into a formula for the generating function $G(x,y,q)$ of Ferrers 
diagrams with a 1-hole
\be
G(x,y,q) = \frac{xy}{(1-qx)(1-qy)} \left( F - x (\partial_x F) - y (\partial_y F) +  
q(\partial_q F) \right)(qx,qy,q), \label{eq:ferrhole}
\ee
where $F(x,y,q)$ is the generating function of Ferrers diagrams (without holes) of height 
and width greater than one.
In particular, we have
\be
G(x,y,1) = \left(\frac{xy}{1-x-y} \right)^3 \frac{1}{(1-x)(1-y)}.
\ee
The perimeter generating function is rational with the same critical point
as the Ferrers diagrams (without holes).
The perimeter and area generating function is given by
\be
G(x,y,q) = \frac{q x^2 y}{(1-qx)(1-qy)} \sum_{n=2}^\infty \frac{(q^2y)^n}{(q^2x;q)_n}
\sum_{k=3}^{n+1}\frac{(k-2)q^k x}{1-q^kx},
\ee
where $(t;q)_n = \prod_{k=0}^{n-1}(1-q^k t)$ denotes the $q$-product.
$G(x,y,q)$ satisfies a $q$-linear equation of third order.
This equation can be derived by expressing $G(x,y,q)$, $G(qx,qy,q)$ and $G(q^2x,q^2y,q)$ in
terms of $F$, $(\partial_x F)$,  $(\partial_y F)$, $(\partial_q F)$ at argument 
$(q^3x,q^3y,q)$.
This is done by using (\ref{eq:ferrhole}) and the symmetrized version of the functional
equation for $F(x,y,q)$.
The resulting system of linear equations can be solved for $F$ in terms of $G$.
Insertion of the result into the functional equation for $F$ gives the functional equation 
for $G$.


\begin{thebibliography}{99}


\bibitem{APP98} Abramov A S, Paule P and Petkov\v sek M 1998 $q$-Hypergemetric Solutions of
$q$-Difference Equations {\it J.~Discr.~Math. \bf 180} 3-22

\bibitem{A31} Adams C R 1931 Linear $q$-difference equations {\it Bull.~Am.Math.~Soc}
{\bf 37} 361-400

\bibitem{B96} Bousquet-M\'elou M 1996 A method for the enumeration of various classes of
column-convex polygons {\it J.~Discr.~Math. \bf 154} 1-25

\bibitem{BM01} Bousquet-M\'elou M 2001 private communication

\bibitem{BO78} Bender C M and Orszag S A 1978 {\it Advanced Mathematical Methods for 
Scientists and Engineers} New York: McGraw-Hill

\bibitem{BOP93} Brak R Owczarek A L and Prellberg T 1993 A scaling theory of the
collapse transition in geometric cluster models {\it J.~Phys.~A \bf 26} 4565-4579

\bibitem{FC82} Fisher M E and Chen J H 1982 Bicriticality and Partial Differential
Approximants, in: {\it Phase Transitions: Cargese 1980}
ed. Levy M Le Guillou J C and Zinn-Justin J New York: Plenum Publishing Corporation
169-216

\bibitem{J00} Janse van Rensburg E J 2000 {\it The Statistical Mechanics of Interacting
walks, Polygons, Animals and Vesicles} New York: Oxford University Press

\bibitem{J001} Janse van Rensburg E J 2000 Interacting columns: generating functions and
scaling exponents {\it J.~Phys.~A \bf 33} 7541-7554

\bibitem{O74} Olver F W J 1974 {\it Asymptotics and Special Functions} New York: 
Academic Press

\bibitem{Ow00} Owczarek A L 2000 private communication

\bibitem{P94} Prellberg T 1994 Uniform $q$-series asymptotics for staircase polygons
{\it J.~Phys.~A \bf 28} 1289-1304

\bibitem{PB95} Prellberg T and Brak R 1995 Critical Exponents from Nonlinear Functional
Equations for Partially Directed Cluster Models {\it J.~Stat.~Phys. \bf 78} 701-730

\bibitem{PO95} Prellberg T and Owczarek A L 1995 Stacking Models of Vesicles and Compact
Clusters {\it J.~Stat.~Phys. \bf 80} 755-779

\bibitem{RF88} Randeria M and Fisher M E 1988 Multisingularity and scaling in partial
differential approximants. I {\it Proc.~R.~Soc.~Lond. A \bf 419} 181-203 

\bibitem{RHHB98} Richard C H\"offe M Hermisson J and Baake M 1998 Random Tilings:
concepts and examples {\it J.~Phys.~A \bf 31} 6385-6408

\bibitem{SF82} Styer D F and Fisher M E 1982 
Partial differential approximants for multivariable power series 
II. Invariance properties {\it Proc.~R.~Soc.~Lond. A \bf 388} 75-102

\bibitem{S90} Styer D F 1990 Subroutine Library for Partial Differential Approximants
{\it Comput. Phys. Commun. \bf 61,3} 374-386 

\bibitem{W89} Wong R 1989 {\it Asymptotic Approximations of Integrals} 
Boston: Academic Press

\end{thebibliography}
\end{document}